\documentclass[aps,showpacs,superscriptaddress,twocolumn]{revtex4}%

\usepackage{amsfonts}
\usepackage{amsmath}
\usepackage{amssymb}
\usepackage{graphicx}%
\begin{document}
\title{Magnetoexcitons and optical absorption of bilayer-structured topological insulators}
\author{Zhigang Wang}
\affiliation{LCP, Institute of Applied Physics and Computational Mathematics, P.O. Box
8009, Beijing 100088, People's Republic of China}
\author{Zhen-Guo Fu}
\affiliation{State Key Laboratory of Superlattices and Microstructures, Institute of
Semiconductors, Chinese Academy of Sciences, P. O. Box 912, Beijing 100083,
People's Republic of China}
\affiliation{LCP, Institute of Applied Physics and Computational Mathematics, P.O. Box
8009, Beijing 100088, People's Republic of China}
\author{Ping Zhang}
\thanks{Corresponding author. Email address: zhang\_ping@iapcm.ac.cn}
\affiliation{LCP, Institute of Applied Physics and Computational Mathematics, P.O. Box
8009, Beijing 100088, People's Republic of China}


\begin{abstract}
The optical absorption properties of magnetoexcitons in topological insulator
bilayers under a strong magnetic field are theoretically studied. A general
analytical formula of optical absorption selection rule is obtained in the
noninteracting as well as Coulomb intra-Landau-level interacting cases, which
remarkably helps to interpret the resonant peaks in absorption spectroscopy
and the corresponding formation of Dirac-type magnetoexcitons. We also discuss
the optical absorption spectroscopy of magnetoexcitons in the presence of
inter-Landau-level Coulomb interaction, which becomes more complex. We hope
our results can be detected in the future magneto-optical experiments.

\end{abstract}
\maketitle

The bilayer $n$-$p$ systems, comprising electrons from the $n$ layer and holes
from the $p$ layer, have been the subject of recent theoretical and
experimental investigations in low-dimensional condensed matter physics. The
coupled quantum wells\ \cite{Snoke,Butov, Timofeev, Eisenstein} and layered
graphene \cite{Zhang,Min,Lozovik1,Iyengar,Berman, Berman1,Koinov} are of
particular interest in connection with the possibility of the Bose-Einstein
condensation and superfluidity of indirect excitons or electron-hole pairs. On
the other hand, as a new state of quantum matter, topological insulators (TIs)
\cite{Kane2005,Bernevig,Fu2007,Konig,Hsieh,Moore,Hasan,Qi,Qi2} are now being
in intensive study. They are characterized by a full insulating gap in the
bulk and topologically protected gapless edge or surface states in low
dimensions, in which by their single-Dirac-point nature it is highly desirable
to find stable and intriguing electron-hole excitations.

\begin{figure}[ptb]
\begin{center}
\includegraphics[width=1.0\linewidth]{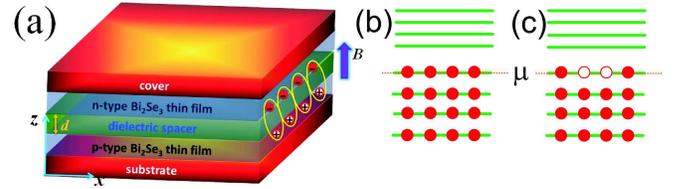}
\end{center}
\caption{(Color online) (a) Scheme of TIB consisting of two Be$_{2}$Se$_{3}$
thin films separated by dielectric spacer. Electron (hole) carriers are
induced by $n$($p$)-type doping or applied external gates. Indirect
magnetoexcitons can be formed in presence of strong external magnetic field.
(b) and (c) are schematic representations of the chemical potential location.
(b) represents the fully occupied case and (c) represents the partially
occupied case.}%
\label{scheme}%
\end{figure}

With the rapid progress in advanced nanotechnology, it is possible to
fabricate a bilayer $n$-$p$ system with TIs. We call this system as
\textquotedblleft topological insulator bilayer (TIB)\textquotedblright, which
comprises two TI thin films separated by a dielectric barrier. The schematic
scheme is shown in Fig. \ref{scheme}(a). Bi$_{2}$Se$_{3}$-family materials are
one of the best candidates for their large bulk-gap width
\cite{Zhang2009,Xia2009,Chen2009}. In order to form steady excitons or
magnetoexcitons, the dielectric spacer can be chosen from those with small
dielectric constant, such as Al$_{2}$O$_{3}$ and SiO$_{2}$, on which,
remarkably, high-quality TI quantum well thin films have been now successively
grown \cite{Chang,Li,Liu,Aguilar}. By doping or applying external gates,
electron and hole carriers will form in both topological insulator thin films,
which behave like massless Dirac particles. However, these quasiparticles can
not form excitons because there is no gap opening. To produce a gap we apply a
strong perpendicular magnetic field on the TIB. In this case, the Dirac-type
energy spectrum of these quasiparticles becomes discrete by forming Landau
levels (LLs), which results in the possible formation of magnetoexcitons in
this system. 

The optical absorption spectroscopy analysis is an instructive method in
studying the properties of magnetoexcitons, because it can provide the
knowledge of magnetoexciton's energies and wavefunctions at $\mathbf{P}$=$0$,
where $\mathbf{P}$ is the magnetoexciton momentum. Many efforts have been
devoted to exploring the optical absorption properties of magnetoexciton in
semiconductor quantum wells by using the magneto-optical measurements
\cite{Marquezini,Kim2004}. More recently, the optical characterization of
Bi$_{2}$Se$_{3}$ in a magnetic field has been experimentally studied
\cite{LaForge2010}. Based on this, the main purpose of the present paper is to
study the magnetoexciton's optical absorption spectroscopy in TIBs. Note that
the magneto-optical response of Bi$_{2}$Se$_{3}$ thin film grown on dielectric
substrate has recently been experimentally measured \cite{Aguilar}. Through
applying a strong magnetic field, the Coulomb interaction (CI) between
electrons in the upper TI film and holes in the lower TI film can be thought
as a perturbation compared to the energy difference between LLS. For
convenience, we divide the CI perturbation into the intralevel [see the
following text and Eq. (\ref{f15})] and interlevel parts. Firstly, we consider
a simple case in which only the intralevel CI is introduced. In this case, we
deduce a general analytical formula for magnetoexciton's optical absorption
selection rule. According to this selection rule, one can clearly point out
which magnetoexciton corresponds to the specific resonant peaks in absorption
spectroscopy. Then, we study the case in which interlevel component of CI is
also included. In this case, the optical absorption spectroscopy becomes more
complex and intriguing by the mixing of the excitations.

We start with the effective Hamiltonian of TIB system $\mathcal{H}$%
=$H_{0}\mathtt{+}U(\mathbf{r})$, where
\begin{equation}
H_{0}=v_{F}\mathbf{\sigma}_{e}\cdot\left(  \mathbf{\hat{z}}\times\mathbf{\pi
}_{e}\right)  -v_{F}\mathbf{\sigma}_{h}\cdot\left(  \mathbf{\hat{z}}%
\times\mathbf{\pi}_{h}\right)  \label{f4}%
\end{equation}
is the free electron-hole part of TIB, while $U(\mathbf{r})$=$-e^{2}%
/\epsilon\sqrt{|\mathbf{r}|^{2}+d^{2}}$ is the CI between the pair of electron
and hole with $\mathbf{r}\mathtt{=}\mathbf{r}_{e}\mathtt{-}\mathbf{r}_{h}$ the
relative coordinate in the $x$-$y$ plane and $d$ the spacer thickness. Here,
$\mathbf{r}_{e(h)}$ represents the position of the electron (hole), and
$\mathbf{\pi}_{e(h)}$ =$\mathbf{p}_{e(h)}\mathtt{\pm}e\mathbf{A}_{e(h)}%
/c$=$-i\partial/\partial\mathbf{r}_{e(h)}\mathtt{\pm}e\mathbf{A}_{e(h)}/c$
denotes the in-plane momentum of the electron (hole), where the gauge is
chosen as $\mathbf{A}_{e(h)}$=$B(0,x_{e(h)},0)$ in the following calculations.
$v_{F}$ is the Fermi velocity ($\mathtt{\sim}3\mathtt{\times}10^{5}$ m/s for
Bi$_{2}$Se$_{3}$-family materials \cite{LiuCX, Wangzg}), $\mathbf{\hat{z}}$ is
the unit vector normal to the surface, and $\mathbf{\sigma}_{e}\mathtt{=}$
$\mathbf{\sigma}\otimes\mathbb{I}$ ($\mathbf{\sigma}_{h}\mathtt{=}$
$\mathbb{I}$ $\mathtt{\otimes}$ $\mathbf{\sigma}$) describes the spin operator
acting on the electron (hole), in which $\mathbf{\sigma}$ denotes a vector of
Pauli matrices and $\mathbb{I}$ is the 2$\times$2 identity matrix.

The eigenstates of Hamiltonian $H_{0}$ have the form
\begin{equation}
\psi_{\mathbf{P}}\left(  \mathbf{R},\mathbf{r}\right)  \mathtt{=}\exp\left[
i\left(  \mathbf{P}\mathtt{+}\frac{e}{2c}\left[  \mathbf{B}\mathtt{\times
}\mathbf{r}\right]  \right)  \mathtt{\cdot}\mathbf{R}\right]  \Psi\left(
\mathbf{r}\mathtt{-}\mathbf{r}_{0}\right)  , \label{f3}%
\end{equation}
where $\mathbf{R}$=($\mathbf{r}_{e}\mathtt{+}\mathbf{r}_{h}$)$/2$,
$\mathbf{r}$=$\mathbf{r}_{e}\mathtt{-}\mathbf{r}_{h}$, and $\mathbf{r}_{0}%
$=$l_{B}^{2}$($\mathbf{\hat{B}}\mathtt{\times}\mathbf{P}$) with $l_{B}$ being
the magnetic length. For an electron in LL $n_{+}$ and a hole in LL $n_{-}$,
the four-component wave functions for the relative coordinate are given by
\cite{Iyengar}%
\begin{align}
&  \Psi_{n_{+},n_{-}}(\mathbf{r})=|n_{+},n_{-}\rangle\label{f13}\\
&  =(\sqrt{2})^{\delta_{n_{+},0}+\delta_{n_{-},0}-2}\left(
\begin{array}
[c]{c}%
\Phi_{|n_{+}|-1,|n_{-}|-1}(\mathbf{r})\\
i^{-\text{sgn}(n_{-})}\Phi_{|n_{+}|-1,|n_{-}|}(\mathbf{r})\\
i^{\text{sgn}(n_{+})}\Phi_{|n_{+}|,|n_{-}|-1}(\mathbf{r})\\
i^{\text{sgn}(n_{+})-\text{sgn}(n_{-})}\Phi_{|n_{+}|,|n_{-}|}(\mathbf{r})
\end{array}
\right)  ,\nonumber
\end{align}
where $\Phi_{n_{1},n_{2}}(\mathbf{r})\mathtt{=}\frac{2^{-\frac{|l_{z}|}{2}%
}n_{\_}!e^{-il_{z}\phi}\delta\left(  l_{z}\right)  r^{|l_{z}|}}{\sqrt{2\pi
n_{1}!n_{2}!}}L_{n_{-}}^{|l_{z}|}(\frac{r^{2}}{2})e^{-\frac{r^{2}}{4}}$ with
$l_{z}$=$n_{1}\mathtt{-}n_{2}$, $n_{-}\mathtt{=}\min(n_{1},n_{2})$, and
$\delta\left(  l_{z}\right)  $=sgn$(l_{z})^{l_{z}}\mathtt{\rightarrow}1$ for
$l_{z}$=$0$. The corresponding LLs are given by
\begin{equation}
E_{n_{+},n_{-}}^{(0)}\mathtt{=}\frac{\sqrt{2}v_{F}}{l_{B}}\mathtt{[}%
\text{sgn}(n_{+})\sqrt{|n_{+}|}\mathtt{-}\text{sgn}(n_{-})\sqrt{|n_{-}%
|}\mathtt{].} \label{f14}%
\end{equation}

Taking CI as a perturbation in the first order and only considering its
intralevel component, the energy dispersion of magnetoexcitons can be easily
obtained as
\begin{equation}
E_{n_{+},n_{-}}\mathtt{=}E_{n_{+},n_{-}}^{(0)}\mathtt{+}\langle\Psi
_{n_{+},n_{-}}|U\left(  \mathbf{r}\mathtt{+}\mathbf{r}_{0}\right)
|\Psi_{n_{+},n_{-}}\rangle. \label{f15}%
\end{equation}
However, to take into account the interlevel CI, we should perform
diagonalization of the full Hamiltonian for Coulomb interacting carriers in
some basis of magnetoexciton states $\Psi_{n_{+},n_{-}}(\mathbf{r})$. To
obtain eigenvalues of the Hamiltonian $\mathcal{H}$, we need to solve the
following equation:%

\begin{align}
0  &  \mathtt{=}\det\left\Vert \delta_{n_{+},n_{+}^{\prime}}\delta
_{n_{-},n_{-}^{\prime}}(E_{n_{+},n_{-}}^{(0)}\mathtt{-}E)\right. \label{f16}\\
&  \left.  \mathtt{+}\langle\Psi_{n_{+}^{\prime},n_{-}^{\prime}}|U\left(
\mathbf{r}\mathtt{-}l_{B}^{2}\mathbf{\hat{z}}\mathtt{\times}\mathbf{P}\right)
|\Psi_{n_{+},n_{-}}\rangle\right\Vert .\nonumber
\end{align}
The location of the chemical potential will determine the possible LL indices
for electrons and holes. For convenience, in this paper we use the notation
$\mu$ for the highest filled LL and only consider two cases: (i) Electron LLs
with $n_{+}$%
%TCIMACRO{\TEXTsymbol{>}}%
%BeginExpansion
$>$%
%EndExpansion
$\mu$ are unoccupied and hole LLs with $n_{-}\mathtt{\leqslant}\mu$ are fully
occupied and (ii) electron LLs with $n_{+}\mathtt{>}\mu$ are unoccupied and
hole LLs with $n_{-}\mathtt{<}\mu$ are fully occupied, while the LL at $\mu$
is partially occupied. These two cases are schematically represented in Fig.
\ref{scheme}(b) and 1(c), respectively. To distinguish these two cases, in the
following discussion we call the first case as fully occupied case and the
second one as partially occupied case. \begin{figure}[ptb]
\begin{center}
\includegraphics[width=1.0\linewidth]{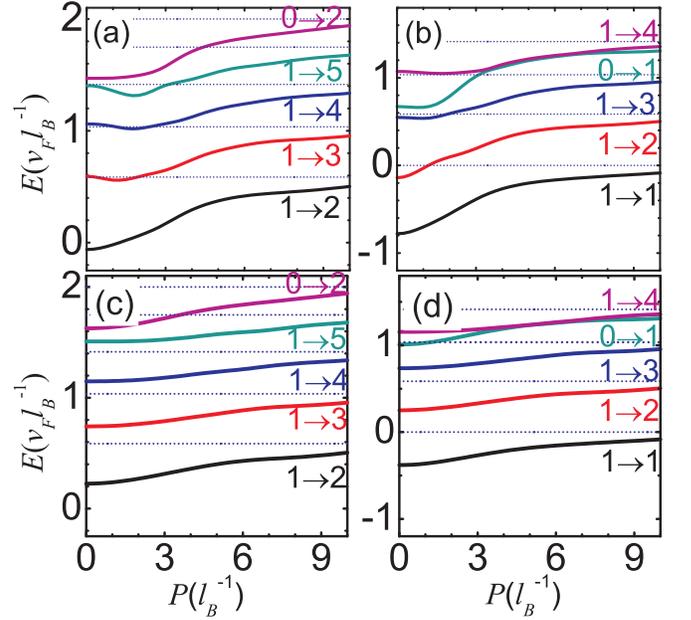}
\end{center}
\caption{(Color online) Energy dispersion of the first five magnetoexciton LLs
for the fully (left panels) and partially (right panels) occupied cases.
$\beta$=$0.9$ and $\mu$=$1$. The spacer thickness $d$=$0.2l_{B}$ for (a) and
(b), while $d$=$2.0l_{B}$ for (c) and (d). }%
\label{figure2}%
\end{figure}

We take $\mu$=$1$ as an example to numerically solve Eq. (\ref{f16}) by
employing five electron and five hole levels. In our calculations we use the
notation $\beta\mathtt{\equiv}\left(  e^{2}/\epsilon l_{B}\right)
/(v_{F}/l_{B})=e^{2}/\left(  \epsilon v_{F}\right)  $ to describe the relative
CI strength. Figures \ref{figure2}(a) and \ref{figure2}(b) plot the first five
energy levels as a function of $P$ for fully and partially occupied cases,
respectively. Here, the spacer thickness is chosen to be $d$=$0.2l_{B}$. With
increasing $d$, the CI effect turns to become faint. As a result, the
magnetoexciton's energy levels become smoother. This can be clearly seen from
Figs. \ref{figure2}(c) and \ref{figure2}(d), where the spacer thickness
$d$=$2.0l_{B}$ is used for comparison with cases of $d$=$0.2l_{B}$.

Let us now turn to discuss the optical absorption properties of
magnetoexcitons with $P=0$. The magneto-optical experiments provides a good
method to detect the magnetoexcitons since the particle-hole excitation energy
could be determined directly by the resonant peaks in the optical
conductivity. Up to this stage, the LLs have been assumed to be sharp, and
thus from Fermi's golden rule, the optical absorption for photons of frequency
$\omega$ yields a sum of delta functions in energy,
\begin{equation}
R(\omega)\text{=}\frac{2\pi}{\hslash}\sum_{\alpha}\left\vert \langle
\alpha|\frac{e}{c}\mathbf{A}\cdot\mathbf{v}|0\rangle\right\vert ^{2}%
\delta(\epsilon_{eh}^{(\alpha)}-\hslash\omega), \label{a1}%
\end{equation}
where $\alpha$ is the set of quantum numbers describing a particle-hole
excitation and $|0\rangle\mathtt{\equiv}|\mu,\mu\rangle$ is the ground state
in absence of particle-hole excitations. $\mathbf{v}$=$\partial\mathcal{H}%
_{0}/\hslash\partial\mathbf{k}$ is the velocity operator for non-interacting
electrons (we neglect the renormalization of the Fermi velocity due to
interactions). $\mathbf{A}$ is the vector potential. For linear polarized
light $\mathbf{A}$=$A\mathbf{\hat{x}}$, one easily deduce $v_{x}%
\mathtt{\varpropto}m_{x}^{-}$=($\sigma_{e}^{y}$\texttt{-}$\sigma_{h}^{y}$)$/2$.

For the purpose of resolving the spatial functional forms and reflecting
realistic experimental conditions one should either assume that the LLs are
broadened or that the magnetoexciton has a finite lifetime. We choose,
therefore, a Lorentzian type broadening $\delta(\epsilon_{eh}^{(\alpha
)}\mathtt{-}\hslash\omega)\mathtt{\rightarrow}\frac{\gamma/2}{(\epsilon
_{eh}^{(\alpha)}-\omega)^{2}+\gamma^{2}/4}$ with linewidth $\gamma$ in
following calculations. Substituting it into Eq. (\ref{a1}), one gets
\begin{equation}
R(\omega)\mathtt{\varpropto}\sum_{\alpha}\left\vert \langle\alpha|m_{x}%
^{-}|0\rangle\right\vert ^{2}\frac{\gamma/2}{(\epsilon_{eh}^{(\alpha
)}\mathtt{-}\omega)^{2}\mathtt{+}\gamma^{2}/4} \label{a4}%
\end{equation}
with $\left\vert \langle\alpha|m_{x}^{-}|0\rangle\right\vert ^{2}%
\mathtt{=}\sum_{n_{+}>\mu}\sum_{n_{-}\leqslant\mu}\int Td\mathbf{r}$ for full
occupation and $\sum_{n_{+}\geqslant\mu}\sum_{n_{-}\leqslant\mu}\int
Td\mathbf{r}$ for partial occupation. The parameter $T$ is defined as%
\begin{equation}
T\mathtt{=}|C_{n_{+},n_{-}}^{\alpha}\langle n_{+},n_{-}|m_{x}^{-}|\mu
,\mu\rangle|^{2}, \label{a6}%
\end{equation}
where $C_{n_{+},n_{-}}^{\alpha}$ is the projection of magnetoexciton state
$\alpha$ on the basis state $|n_{+},n_{-}\rangle$.

Let us now consider a simple case in which only the intralevel CI is
considered. This case is similar to that without CI, in which the
magnetoexciton state $\alpha$ is one special basis state and $C_{n_{_{+}%
},n_{_{-}}}^{\alpha}$=$\delta_{\alpha,(n_{_{+}},n_{_{-}})}$. So by
substituting Eq. (\ref{f13}) into $\langle n_{+},n_{-}|m_{x}^{-}|\mu
,\mu\rangle$, we can immediately obtain the optical absorption selection rule
for magnetoexcitons as
\begin{align}
&  \langle n_{+},n_{-}|m_{x}^{-}|\mu,\mu\rangle\label{a7}\\
&  \mathtt{=}c_{1}\delta_{|n_{+}|,|\mu|}\delta_{|n_{-}|,|\mu|-1}%
\mathtt{+}c_{2}\delta_{|n_{+}|,|\mu|-1}\delta_{|n_{-}|,|\mu|}\nonumber\\
&  \mathtt{+}c_{3}\delta_{|n_{+}|,|\mu|}\delta_{|n_{-}|,|\mu|+1}%
\mathtt{+}c_{4}\delta_{|n_{+}|,|\mu|+1}\delta_{|n_{-}|,|\mu|},\nonumber
\end{align}
where the coefficients $c_{i}$ ($i$=$1$, $\cdots$, $4$) are defined as%
\begin{align}
c_{1}  &  \mathtt{=}\left[  i(-i)^{-\text{sgn}(n_{-})}\mathtt{+}%
i^{1+\text{sgn}(\mu)}(-i)^{\text{sgn}(n_{+})-\text{sgn}(n_{-})}\right]
C,\text{ }\nonumber\\
c_{2}  &  \mathtt{=-}\left[  i(-i)^{\text{sgn}(n_{+})}\mathtt{+}%
i^{1-\text{sgn}(\mu)}(-i)^{\text{sgn}(n_{+})-\text{sgn}(n_{-})}\right]
C,\nonumber\\
c_{3}  &  \mathtt{=}\left[  (-i)^{1-\text{sgn}(\mu)}\mathtt{+}%
(-i)^{1+\text{sgn}(n_{+})}\right]  C,\text{ }\nonumber\\
c_{4}  &  \mathtt{=}\left[  i^{1+\text{sgn}(\mu)}\mathtt{+}i(-i)^{-\text{sgn}%
(n_{-})}\right]  C, \label{a81}%
\end{align}
with $C\mathtt{=}(\sqrt{2})^{\delta_{n_{+},0}+\delta_{n_{-},0}+2\delta_{\mu
,0}-4}$. Our optical absorption selection rule formula (\ref{a7}) proves to be
very useful when analyzing magnetoexciton's absorption resonant peaks in
spectroscopy of TIB systems. For the fully occupied case with $\mu
\mathtt{\geqslant}0$, the selection rule can further be simplified as
\begin{equation}
\langle n_{+},n_{-}|m_{x}^{-}|\mu,\mu\rangle=c_{4}\delta_{n_{+},\mu+1}%
\delta_{n_{-},\mu}. \label{a10}%
\end{equation}
\begin{figure}[ptb]
\begin{center}
\includegraphics[width=1.0\linewidth]{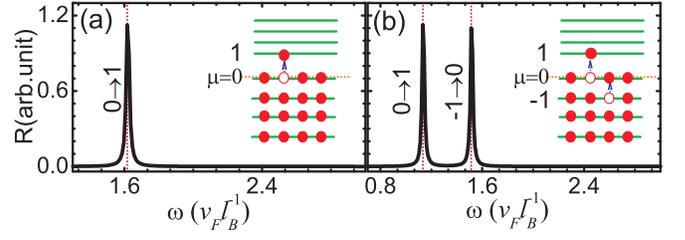}
\end{center}
\caption{{}(Color online) The optical absorption curves with $\mu$=$0$ for (a)
fully and (b) partially occupied cases. The spacer thickness is set to
$d$=$0.2l_{B}$. The corresponding schematic sketch of magnetoexcitons are
plotted in the insets.}%
\label{figure4}%
\end{figure}The present general optical absorption selection rule proves to be
very useful when analyzing magnetoexciton's absorption resonant peaks in
spectroscopy of TIB systems.

For simplicity, let us first consider the fully occupied case with $\mu$=$0$,
i.e., electron LLs $n_{+}$%
%TCIMACRO{\TEXTsymbol{>}}%
%BeginExpansion
$>$%
%EndExpansion
$0$ are unoccupied and hole LLs $n_{-}\mathtt{\leqslant}0$ are fully occupied.
The ground state is $|0\rangle$=$|0,0\rangle$. According to the selection rule
(\ref{a10}), the non-zero elements of $\langle n_{+},n_{-}|m_{x}%
^{-}|0,0\rangle$ only occurs at $n_{+}$=$1$ while $n_{-}$=$0$. That means
there is only one resonant peak to appear in the absorption spectrum, which
corresponds to the energy of magnetoexciton state $|1,0\rangle$ (see Fig.
\ref{figure4}(a)). For comparison, we also study the partially occupied case
with $\mu$=$0$. According to the selection rule (\ref{a7}), $\langle
n_{+},n_{-}|m_{x}^{-}|0,0\rangle\mathtt{\neq}0$ only when $n_{+}$=$1$, $n_{-}%
$=$0$ or $n_{+}$=$0$, $n_{-}$=$-1$. Thus in this case there are \emph{two}
resonant peaks occurring in the absorption spectrum as shown in (see Fig.
\ref{figure4}(b)), which correspond to the energies of two magnetoexciton
states $|1,0\rangle$ and $|0,-1\rangle$. By increasing $d$ (equally decreasing
the CI), these two magnetoexciton levels $|1,0\rangle$ and $|0,-1\rangle$ turn
to be degenerate, which results in the incorporation of these two resonant
peaks to be a single one at $d\mathtt{\rightarrow}\infty$. Schematic sketch of
the corresponding magnetoexciton state obeying the selection rule (\ref{a10})
are plotted in the insets of Fig. \ref{figure4}. \begin{figure}[ptb]
\begin{center}
\includegraphics[width=1.0\linewidth]{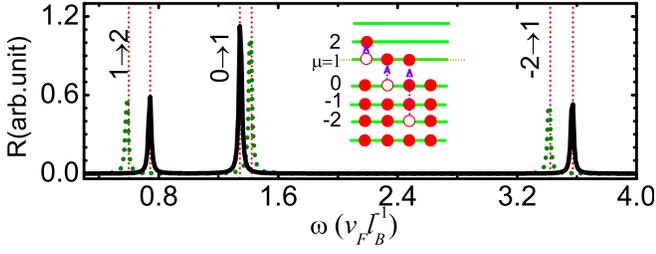}
\end{center}
\caption{{}(Color online) Same as Fig. \ref{figure4}(b) except $\mu$=$1$. The
green dotted curve represents the corresponding optical absorption spectrum at
the limit of $d\mathtt{\rightarrow}\infty$.}%
\label{figure5}%
\end{figure}

To ulteriorly see the difference of absorption properties between partially
and fully occupied cases, let us now consider the case of $\mu$=$1$, i.e., the
ground state is $|1,1\rangle$. The selection rule for fully occupied case
(Eq.(\ref{a10})) with $\mu$=$1$ promises that there is only one resonant peak
occurs, which corresponds to the formation of magnetoexciton state
$|2,1\rangle$ by absorbing a photon quanta $\omega$. In the limit
$d\mathtt{\rightarrow}\infty,$ $\omega\mathtt{\rightarrow}\sqrt{2}(\sqrt
{2}\mathtt{-}1)$=$0.59$. For partially occupied case, however, there are three
non-zero elements according to selection rule (\ref{a7}), which respectively
are $n_{+}$=$1$, $n_{-}$=$0$; $n_{+}$=$1$, $n_{-}$=$-2$; and $n_{+}$=$2$,
$n_{-}$=$1$. So there are \emph{three} resonant peaks appearing in optical
absorption spectrum (see black curves in Fig. \ref{figure5} for $d$=$0.2l_{B}%
$), which correspond to the formation of magnetoexciton states $|2,1\rangle$,
$|1,0\rangle$, and $|1,-2\rangle$, respectively, by absorbing photons of three
different frequencies. As increasing the spacer thickness, these three
absorption frequencies $\omega_{1}\mathtt{\rightarrow}\sqrt{2}(\sqrt
{1}\mathtt{-}0)$=$1.41$, $\omega_{2}\mathtt{\rightarrow}\sqrt{2}%
(1\mathtt{+}\sqrt{2})$=$3.41$, and $\omega_{3}\mathtt{\rightarrow}\sqrt
{2}(\sqrt{2}\mathtt{-}1)$=$0.59$ (see green dotted curves in Fig. 4).

\begin{figure}[ptb]
\begin{center}
\includegraphics[width=1.0\linewidth]{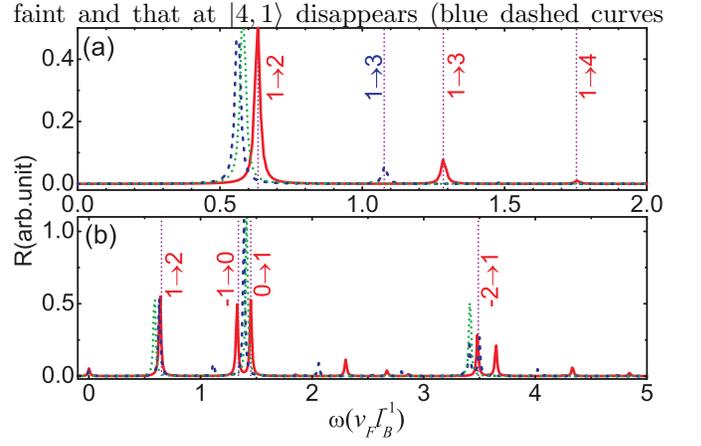}
\end{center}
\caption{{}(Color online) The optical absorption spectrums with $\mu$=$1$ for
(a) fully and (b) partially occupied cases when the interlevel Coulomb
interaction is included. The red, blue and green curves are for $d$=$0.2l_{B}%
$, $2.0l_{B}$, and $10.0l_{B}$, respectively.}%
\label{figure6}%
\end{figure}

When the interlevel CI is also included, the above-discussed magnetoexciton
states will be mixed and energetically redistributed, which observably correct
the exact selection rule (\ref{a7}) and bring about additional subsidiary
peaks in the optical absorption spectrum. To clearly see this, we reconsider
the case of $\mu$=$1$ as an example. The corresponding calculated optical
absorption spectra for fully and partially occupied cases are respectively
shown in Fig. \ref{figure6}(a) and \ref{figure6}(b). 

Let us first make an analysis on the fully occupied case. Because the
interlevel CI mixes the noninteracting LLs, the absorption phenomenon should
appear at the energy of every magnetoexciton state, with the resultant
absorption peak amplitude decided by the projection of the corresponding
resonant excited states $|n_{+},n_{-}\rangle$ allowed by the selection rule
(\ref{a7}) in the case without interlevel interaction. When the spacer
thickness $d$ is small (e.g. $d$=$0.2l_{B}$), not only the main resonant peak
appears at the energy of excited level $|2,1\rangle$, but also two weak
resonant peaks at the energies of excited levels $|3,1\rangle$ and
$|4,1\rangle$ can be observed (see red curves in Fig. \ref{figure6}(a)). By
increasing $d$ to $2.0l_{B}$, the resonant peak at $|3,1\rangle$ become faint
and that at $|4,1\rangle$ disappears (blue dashed curves in Fig.
\ref{figure6}(a)). When further increasing $d$ to $10.0l_{B}$, we find that
the resonant peak $|3,1\rangle$ also disappears (green dotted curves in Fig.
\ref{figure6}(a)) and the optical absorption spectrum comes back to the
noninteracting case.

Similar physics also happens for partially occupied case (see Fig.
\ref{figure6}(b)). The overlap caused by the interlevel CI between
magnetoexciton states will produce many subsidiary resonant peaks besides the
main peaks at the energies of magnetoexciton states $|2,1\rangle$,
$|1,0\rangle$ and $|1,-2\rangle$. Note that there is a subsidiary resonant
peak appearing in front of $|2,1\rangle$, which corresponds to the ground
state $|1,1\rangle$. This special feature only appears for the partially
occupied case and thus can be used to experimentally determine whether the
highest occupied LL is fully filled or not in TIB. By increasing $d$, the
subsidiary resonant peaks turn to disappear, and the optical absorption
spectrum turns back to the noninteracting case: only the main three resonant
peaks at energies of states $|2,1\rangle$, $|1,0\rangle$ and $|1,-2\rangle$
appear in absorption spectrum allowed by the selection rule (\ref{a7}).

In summary, we have theoretically studied the optical absorption properties of
magnetoexcitons formed in TIB under a strong perpendicular magnetic field. By
neglecting inter-Landau-level CI, we have determined an exact absorption
selection rule, which greatly helps to connect absorption peaks with specific
magnetoexcitons. The inclusion of inter-Landau-level CI has been shown to
bring about subsidiary optical absorption peaks due to the mixing effect. Our
present conclusion also works for graphene bilayers.

This work was supported by NSFC under Grants No. 10904005 and No. 90921003,
and by the National Basic Research Program of China (973 Program) under Grant
No. 2009CB929103.


\begin{thebibliography}{99}                                                                                               %


\bibitem {Snoke}D.W. Snoke, Science \textbf{298}, 1368 (2002).

\bibitem {Butov}L.V. Butov, J. Phys.: Condens. Matter \textbf{16}, R1577 (2004).

\bibitem {Timofeev}V.B. Timofeev and A.V. Gorbunov, J. Appl. Phys.
\textbf{101}, 081708 (2007).

\bibitem {Eisenstein}J.P. Eisenstein and A.H. MacDonald, Nature (London)
\textbf{432}, 691 (2004).

\bibitem {Zhang}C.H. Zhang and Y.N. Joglekar, Phys. Rev. B \textbf{77}, 233405 (2008).

\bibitem {Min}H. Min, R. Bistritzer, J.-J. Su, and A. H. MacDonald, Phys. Rev.
B \textbf{78}, 121401(R) (2008).

\bibitem {Lozovik1}Yu. E. Lozovik and A. A. Sokolik, Pis'ma v ZhETF,
\textbf{87}, 61 (2008). [JETP Lett. \textbf{87}, 55 (2008)].

\bibitem {Iyengar}A. Iyengar, J.H. Wang, H.A. Fertig, and L. Brey, Phys. Rev.
B \textbf{75}, 125430 (2007).

\bibitem {Berman}O.L. Berman, Y.E. Lozovik, and G. Gumbs, Phys. Rev. B
\textbf{77}, 155433 (2008).

\bibitem {Berman1}O.L. Berman, R.Ya. Kezerashvili, and Y.E. Lozovik, Phys.
Rev. B \textbf{78}, 035135 (2008).

\bibitem {Koinov}Z.G. Koinov, Phys. Rev. B \textbf{79}, 073409 (2009).

\bibitem {Kane2005}C.L. Kane and E.J. Mele, Phys. Rev. Lett. \textbf{95},
226801 (2005); \textbf{95}, 146802 (2005).

\bibitem {Bernevig}B.A. Bernevig and S.C. Zhang, Phys. Rev. Lett. \textbf{96},
106802 (2006).

\bibitem {Fu2007}L. Fu and C. L. Kane, Phys. Rev. B \textbf{76}, 045302 (2007).

\bibitem {Konig}M. K\"{o}nig, S. Wiedmann, Christoph Br\"{u}ne, A. Roth, H.
Buhmann, L. W. Molenkamp, X.-L. Qi, and S. C. Zhang, Science \textbf{318}, 766 (2007).

\bibitem {Hsieh}D. Hsieh, D. Qian, L. Wray, Y. Xia, Y. S. Hor, R. J. Cava, and
M. Z. Hasan, Nature \textbf{452}, 970 (2008).

\bibitem {Moore}J.E. Moore, Nature (London) \textbf{464}, 194 (2010).

\bibitem {Qi}X.-L. Qi and S.-C. Zhang, Phys. Today \textbf{63}, 33 (2010).

\bibitem {Hasan}M.Z. Hasan and C.L. Kane, Rev. Mod. Phys. \textbf{82}, 3045
(2010), and references therein.

\bibitem {Qi2}X.L. Qi and S.C. Zhang, Rev. Mod. Phys. \textbf{83}, 1057
(2011), and references therein.

\bibitem {Zhang2009}H.-J. Zhang, C.-X. Liu, X.-L. Qi, X. Dai, Z. Fang, and
S.-C. Zhang, Nature Phys. \textbf{5}, 438 (2009).

\bibitem {Xia2009}Y. Xia, D. Qian, D. Hsieh, L. Wrayl, A. Pal1, H. Lin, A.
Bansil, D. Grauer, Y. S. Hor, R. J. Cava, Nat. Phys. \textbf{5}, 398 (2009).

\bibitem {Chen2009}Y. L. Chen, J. G. Analytis, J. H. Chu, Z. K. Liu, S. K. Mo,
X. L. Qi, H. J. Zhang, D. H. Lu, X. Dai, Z. Fang, Science \textbf{325}, 178 (2009).

\bibitem {Chang}C.-Z. Chang, K. He, L.-L. Wang, X.-C. Ma, M.-H. Liu, Z.-C.
Zhang, X. Chen, Y.-Y. Wang, and Q.-K. Xue, Spin \textbf{1}, 21 (2011).

\bibitem {Li}Z. Li, Y. Qin, Y. Mu, T. Chen, C. Xu, L. He, J. Wan, F. Song, M.
Han, G. Wang, J. Nanosci. Nanotechnol. \textbf{11}, 7042 (2011).

\bibitem {Liu}H. Liu and P.D. Ye, Appl. Phys. Lett. \textbf{99}, 052108 (2011).

\bibitem {Aguilar}R. Vald\'{e}s Aguilar, A. V. Stier, W. Liu, L.S. Bilbro, D.
K. George, N. Bansal, L. Wu, J. Cerne, A. G. Markelz, S. Oh, and N.P.
Armitage, Phys. Rev. Lett. \textbf{108}, 087403 (2012).

\bibitem {Marquezini}M. V. Marquezini, M. J. S. P. Brasil, M. A. Cotta, and J.
A. Brum, Phys. Rev. B \textbf{53}, R16156 (1996).

\bibitem {Kim2004}Y. Kim, K.-S. Lee, and C. H. Perry, Appl. Phys. Lett.
\textbf{84}, 738 (2004).

\bibitem {LaForge2010}A. D. LaForge, A. Frenzel, B. C. Pursley, T. Lin, X.
Liu, J. Shi, and D. N. Basov, Phys. Rev. B \textbf{81}, 125120 (2010).

\bibitem {LiuCX}C.-X. Liu, X.-L. Qi, H. Zhang, X. Dai, Z. Fang, and S.-C.
Zhang, Phys. Rev. B \textbf{82}, 045122 (2010).

\bibitem {Wangzg}Z. Wang, Z.-G. Fu, S.-X. Wang, and P. Zhang, Phys. Rev. B
\textbf{82}, 085429 (2010).
\end{thebibliography}
\end{document}